\documentclass[conference]{IEEEtran}

\usepackage{times}
\usepackage{helvet}
\usepackage{courier}
\usepackage{textcomp}
\usepackage{multirow}
\usepackage{amsmath,amssymb,amsfonts}
\usepackage{algorithmic}
\usepackage{graphicx}
\usepackage{textcomp}
\usepackage{xcolor}
\usepackage{stfloats}
\usepackage{gensymb}
\usepackage{subfigure}
\usepackage{booktabs}
\usepackage[ruled,vlined]{algorithm2e}
\usepackage{makecell}
\usepackage{enumerate}
\newcolumntype{C}[1]{>{\centering\let\newline\\\arraybackslash\hspace{0pt}}m{#1}}
\usepackage{comment}

\usepackage{listings}

\usepackage{xcolor}

\definecolor{codegreen}{rgb}{0,0.6,0}
\definecolor{codegray}{rgb}{0.5,0.5,0.5}
\definecolor{codepurple}{rgb}{0.58,0,0.82}
\definecolor{backcolour}{rgb}{0.95,0.95,0.92}

\lstdefinestyle{mystyle}{
  commentstyle=\color{codegreen},
  keywordstyle=\color{magenta},
  numberstyle=\tiny\color{codegray},
  stringstyle=\color{codepurple},
  basicstyle=\ttfamily\footnotesize,
  breakatwhitespace=false,         
  breaklines=true,                 
  captionpos=b,                    
  keepspaces=true,                 
  showspaces=false,                
  showstringspaces=false,
  showtabs=false,                  
  tabsize=2
}

\begin{document}
%
\title{EdgeFL: A Lightweight Decentralized Federated Learning Framework}



\author{
    \IEEEauthorblockN{Hongyi Zhang\IEEEauthorrefmark{1}, Jan Bosch\IEEEauthorrefmark{1}, Helena Holmström Olsson\IEEEauthorrefmark{2} }
    \IEEEauthorblockA{\IEEEauthorrefmark{1}\textit{Chalmers University of Technology}, Gothenburg, Sweden. Email: \{hongyiz, jan.bosch\}@chalmers.se}

    \IEEEauthorblockA{\IEEEauthorrefmark{2}\textit{Malmö University}, Malmö, Sweden. Email: helena.holmstrom.olsson@mau.se}

}


\maketitle

\begin{abstract}

Federated Learning (FL) has emerged as a promising approach for collaborative machine learning, addressing data privacy concerns. However, existing FL platforms and frameworks often present challenges for software engineers in terms of complexity, limited customization options, and scalability limitations.  In this paper, we introduce EdgeFL, an edge-only lightweight decentralized FL framework, designed to overcome the limitations of centralized aggregation and scalability in FL deployments. By adopting an edge-only model training and aggregation approach, EdgeFL eliminates the need for a central server, enabling seamless scalability across diverse use cases. With a straightforward integration process requiring just four lines of code (LOC), software engineers can easily incorporate FL functionalities into their AI products. Furthermore, EdgeFL offers the flexibility to customize aggregation functions, empowering engineers to adapt them to specific needs. Based on the results, we demonstrate that EdgeFL achieves superior performance compared to existing FL platforms/frameworks. Our results show that EdgeFL reduces weights update latency and enables faster model evolution, enhancing the efficiency of edge devices. Moreover, EdgeFL exhibits improved classification accuracy compared to traditional centralized FL approaches. By leveraging EdgeFL, software engineers can harness the benefits of federated learning while overcoming the challenges associated with existing FL platforms/frameworks.

\end{abstract}


\section{Introduction}

Federated learning is a machine learning approach that enables training models on decentralized data sources while preserving data privacy. Traditional machine learning models typically require centralizing all the data in one location for training, which can be challenging due to data privacy concerns, legal restrictions, and computational constraints. Federated learning addresses these limitations by allowing models to be trained directly on the devices or servers where the data is generated or stored, without the need to transfer it to a central location \cite{mlchallenge}.

The key advantage of federated learning is its ability to preserve data privacy. Since the data remains on the client devices or servers, it alleviates concerns related to data exposure or sharing sensitive information. Instead of sharing raw data, only the updates to the model parameters, often encrypted or anonymized, are communicated during the training process. This decentralized approach enables organizations and individuals to collaborate on model training without compromising the privacy and security of their data.

Federated learning finds applications in various domains, such as healthcare, finance, Internet of Things (IoT), and edge computing \cite{lu2019collaborative}\cite{hu2018federated}\cite{brisimi2018federated}. It allows organizations to leverage the collective knowledge of distributed datasets without violating privacy regulations or exposing sensitive information. Additionally, federated learning can reduce the reliance on costly data transfers and enable training on resource-constrained devices.

Despite the benefits of Federated Learning, Federated learning presents unique challenges for software engineers involved in AI engineering \cite{lwakatare2019taxonomy}. They need to design and develop distributed systems that can effectively handle the communication, coordination, and synchronization between multiple clients and a central server. This requires scalable and fault-tolerant systems to accommodate large-scale deployments while optimizing network utilization and ensuring data consistency across distributed nodes. 

In addition, algorithmic development and optimization are essential for efficient federated learning. Software engineers need to understand and implement federated learning algorithms and optimization techniques, including Federated Averaging and adaptive learning rate strategies \cite{towards}. Developing algorithms that converge to high-quality models while minimizing resource consumption is a challenging task for software engineers. 

Integration with edge devices and IoT environments adds another layer of complexity. Software engineers must consider the resource constraints of these devices, such as limited computational power and energy consumption \cite{rodrigues2019machine}. They need to optimize models and algorithms to fit within these constraints and devise efficient mechanisms for model deployment, updates, and synchronization on edge devices.

Moreover, software engineers may face challenges related to the availability of comprehensive tooling and frameworks for federated learning. They may need to adapt existing tools, develop custom solutions, or contribute to open-source projects to address specific needs. Building efficient workflows, debugging mechanisms, and monitoring tools tailored to federated learning systems is a demanding task.

The existing federated learning (FL) platforms and frameworks from both academia and industry are usually complex to use. FL involves distributed systems, optimization algorithms, privacy techniques, and machine learning models. Integrating these components into a cohesive platform or framework can result in intricate systems with numerous dependencies and configurations. Understanding and navigating these technical intricacies can be challenging for users, especially those without a strong background in distributed systems or machine learning. In addition, FL is applied across a wide range of domains and use cases, each with its specific requirements and constraints. Designing a platform or framework that accommodates this heterogeneity can lead to increased complexity. It becomes a challenge to strike a balance between providing flexibility for customization and maintaining simplicity for ease of use. Last but not least, most of the existing FL frameworks use centralized aggregation in federated learning application development. Relying on a central server creates problems when deploying into production, such as single point of failure, large communication overhead, lack of scalability, etc. In this paper, we present EdgeFL, a lightweight federated learning (FL) framework designed specifically for edge computing environments, aiming to address the challenges associated with centralized aggregation. By adopting an edge-only model training and aggregation approach, EdgeFL eliminates the need for a central server, thereby enabling seamless scalability across diverse use cases. The framework offers a straightforward integration process, requiring only four lines of code (LOC) for software engineers to incorporate FL functionalities into their AI products. Moreover, EdgeFL facilitates the customization of aggregation functions, empowering engineers to customize according to their needs. Thus, the contributions of the paper are the following:

\begin{table*}[b]
\caption{Comparison between existing FL platforms/frameworks and our proposed EdgeFL}
\label{tab:compareexist}
\begin{center}
\begin{tabular}{c c c c c c | c}

\toprule
 & TFF & PySyft  &  FATE  & LEAF & PaddleFL & \makecell{EdgeFL\footnotemark[1]\\(Proposed Framework)} \\

\midrule
Line of Code (LOC) for FL Feature &  $\thicksim$50 &  $\thicksim$150 &  $\thicksim$100 &  $\thicksim$350 &  $\thicksim$100 & \textbf{4}\\ 
Centralized Server required & Yes & Yes & Yes & Yes & Yes & \textbf{No}\\
Asynchronous Node Join & No & No & No & No & No & \textbf{Yes} \\
Heterogeneous Environment Support & No & No & No & No & Limited & \textbf{Yes}\\
Customizable Aggregation & No & No & No & No & No & \textbf{Yes} \\
Containerized Edge Deployability & No & No & Limited & No & Limited & \textbf{Yes}\\

\bottomrule

\end{tabular}
\end{center}
\end{table*}

\noindent 1). We introduce EdgeFL\footnote{https://github.com/HarryME-zh/EdgeFL.git}, the first scalable edge-only FL framework. To accomplish easy-implementation and scalable model training capacity, simple API design and learning flow abstraction are built.

\noindent 2). To expedite industrial FL training and support large-scale node communication among edges, we suggested a decentralized FL architecture and learning algorithm which enables asynchronous model training. The architecture could serve as a model for future edge-only FL development and study.

\noindent 3). Based on flexible user definitions, engineers and researchers can quickly construct a customizable model and aggregation approach for diverse needs.

\noindent 4). EdgeFL offers scalable and seamless deployment capabilities to facilitate rapid prototyping and production-level training for both industrial and research purposes.

The remainder of this paper is structured as follows. 
In Section II, we introduce the background and related work of this study. Section III details our research method, including the implementation, data distribution, machine learning methods applied and evaluation metrics. Section IV presents the system design of our proposed EdgeFL.  Section V evaluates our proposed framework and compared it with existing Federated Learning frameworks/platforms. Section VI outlines the discussion on our observed results. Finally, Section VII presents conclusions and future work.

\section{Background and Related Work}
\label{sec: back}

\subsection{Federated Learning}

Federated learning (FL) is a machine learning paradigm that allows multiple decentralized devices or entities to collaboratively train a shared model without sharing their raw data \cite{mcmahan2017communication}\cite{li2020review}\cite{zhang2021survey}. In traditional machine learning approaches, a central server or data aggregator collects and stores all the training data from different sources. The central server then trains a global model using the combined data. However, this centralized approach raises concerns regarding data privacy, security, and the practicality of transferring large volumes of data to a central location \cite{towards}. With the rise of edge computing and distributed data sources, there is a growing need to leverage data that resides on various devices or entities while respecting data ownership and privacy. Federated learning addresses this challenge by allowing local devices, such as smartphones, IoT devices, or edge servers, to train a shared model using their locally stored data. As shown in Figure \ref{fig:fl-diag}, the FL process typically involves the following steps:

\begin{figure}[!htpb]
\centerline{\includegraphics[scale=0.32]{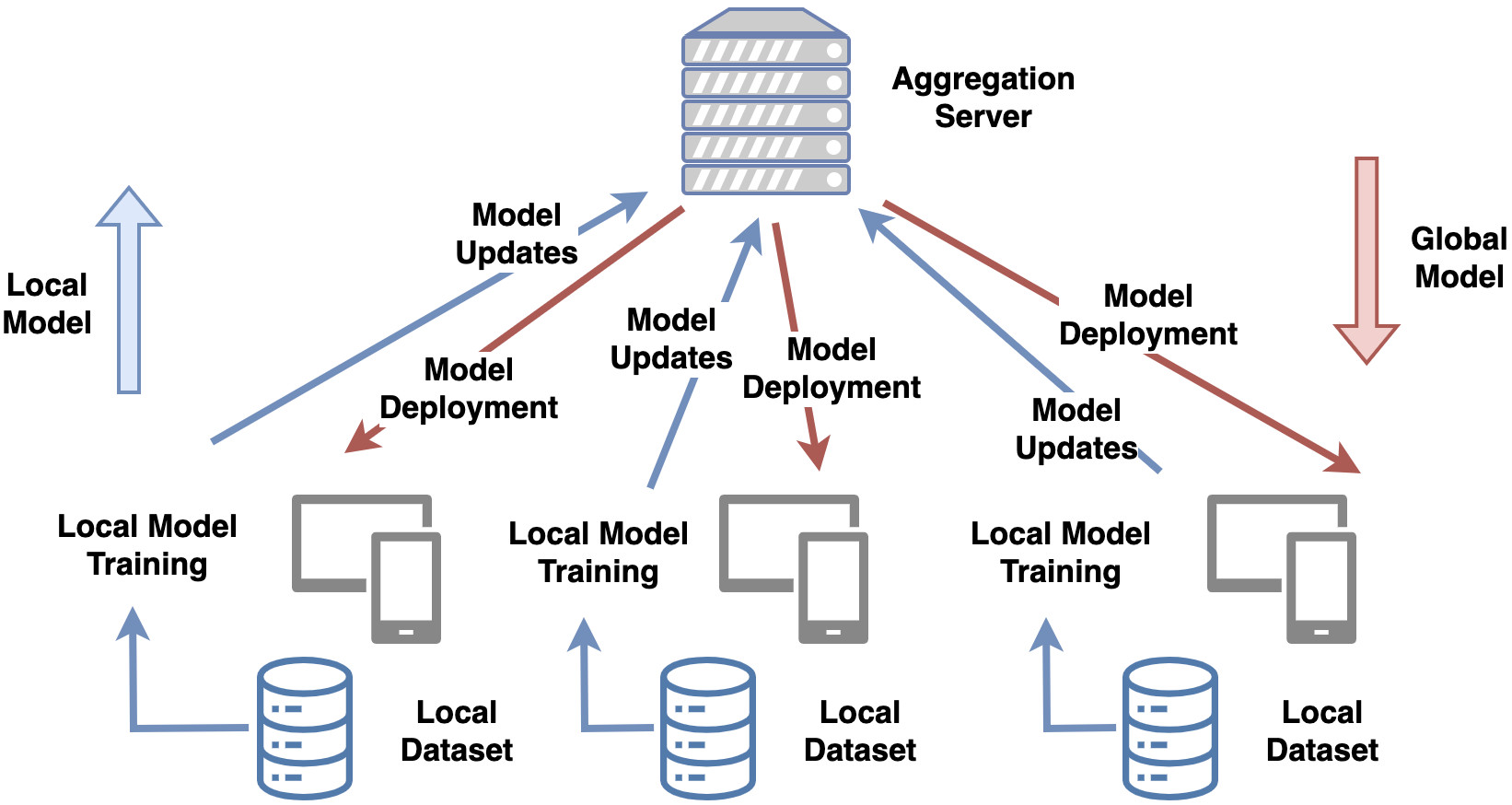}}

\caption{Diagram of Federated Learning training process}
\label{fig:fl-diag}
\end{figure}

1). Initialization: The central server initializes a global model and distributes it to the participating devices or entities.

2). Local Model Training: Each device or entity independently trains the global model using its local data, without sharing the raw data. The local model is trained using gradient descent or other optimization algorithms, updating the model parameters based on its local data.

3). Model Aggregation: After local training, the devices or entities send their locally computed model updates (such as gradients) to the central server. The central server aggregates these updates using techniques like Federated Averaging or Secure Aggregation to obtain a refined global model.

4). Iterative Process: The model training and aggregation process iterates over multiple rounds, allowing the global model to improve over time. Each round typically consists of local model training, model aggregation, and communication between devices and the central server.

FL enables collaborative learning from a diverse range of devices or entities, each having its unique data distribution and characteristics. This diversity helps improve the generalization and robustness of the global model by capturing a more comprehensive representation of the data.

\subsection{Existing Federated Learning Platform/Framework}

There are several existing federated learning (FL) platforms and frameworks available that facilitate the development and deployment of FL systems. TensorFlow Federated is an open-source research-oriented framework developed by Google. It extends TensorFlow with FL capabilities, enabling the implementation of FL algorithms and protocols. TFF provides a programming model and APIs for expressing federated computations, along with tools for simulation and evaluation of FL algorithms \cite{tensorflow2021tensorflow}. PySyft is an open-source library built on top of PyTorch that focuses on privacy-preserving FL and secure multi-party computation \cite{ziller2021pysyft}. It provides a high-level API for expressing FL computations and offers privacy techniques like differential privacy and secure aggregation. It allows users to work with remote data and models, enabling collaborative learning while protecting privacy. FATE is an industrial-grade FL framework developed by Webank Research \cite{liu2021fate}. It provides a secure and flexible infrastructure for collaborative model training across distributed entities while preserving data privacy. FATE supports a wide range of machine learning algorithms and provides functionalities for data preprocessing, model evaluation, and privacy protection. It has been widely adopted in industries such as finance, healthcare, and telecommunications. LEAF is an open-source FL framework developed by NVIDIA \cite{caldas2018leaf}. It offers a comprehensive set of tools and functionalities for researchers and practitioners to experiment with and evaluate FL algorithms. PaddleFL is a FL framework developed by PaddlePaddle, an open-source deep-learning platform \cite{ma2019paddlepaddle}. It provides a distributed infrastructure for training large-scale FL models. PaddleFL supports various optimization algorithms, model architectures, and communication protocols. 

However, the current FL platforms and frameworks available in academia and industry are complex, demanding a thorough understanding of FL concepts. As shown in Table \ref{tab:compareexist}, the majority of existing FL systems require the implementation of more than 100 lines of code (LOC) to deploy a FL application, providing limited flexibility for aggregation function customization and lacking support for asynchronous communication schemes. These constraints present significant challenges to software engineers attempting to integrate FL into production environments. Furthermore, because all of these platforms/frameworks rely on a centralized aggregation server, issues such as single-point failure and scalability constraints arise.

\section{Research Method}

In this research, we adopted the empirical methodology and learning procedure outlined by Zhang \cite{zhang2003machine} to conduct a comprehensive quantitative measurement and evaluation of our proposed EdgeFL framework in comparison to existing Federated Learning platforms/frameworks. We aim to present a thorough and robust assessment of the performance and effectiveness of the EdgeFL framework. In the subsequent sections, we provide detailed insights into our implementation approach, present the method employed for dataset partitioning and distribution for heterogeneous simulation, discuss the evaluation metrics employed, and elaborate on the machine learning methods utilized during the experimental analysis.

\subsection{Implementation}

In order to thoroughly evaluate the performance and capabilities of the EdgeFL framework, we conducted experiments using two widely recognized machine learning applications: digit recognition and object recognition. For these experiments, we leveraged the MNIST and CIFAR-10 datasets, which are extensively used in the research field. To facilitate deep learning training and testing, we developed the applications using the PyTorch backend. With the integration of EdgeFL, the FL functionality is seamlessly integrated into these machine learning applications with the addition of just four lines of code (LOC). Furthermore, we containerized the applications, enabling easy deployment on edge devices while maintaining their functionality and performance.

The flexibility of the EdgeFL framework allows it to be constructed on various container orchestration clusters, such as Kubernetes, Docker Swarm, etc. In this paper, we utilized Docker Swarm \cite{soppelsa2016native} as the cluster of choice. Docker Swarm offers an efficient and scalable environment for managing containerized applications. The services within Docker Swarm facilitate seamless communication among containers, while an internal DNS resolver ensures peer node service communication. By utilizing the capabilities of Docker Swarm, we were able to create a robust and scalable deployment environment for the EdgeFL framework, ensuring its suitability for edge devices and distributed computing scenarios.

\subsection{Dataset Distribution}
\label{sec:distribution}

For the purpose of this study, we used two kinds of edge data distribution to analyze system performance for heterogeneous simulation. 

\subsubsection {Uniform Distribution}

Within this experimental setup, the training data samples were distributed among the edge nodes following a uniform distribution. This distribution ensured an equal likelihood of data samples from each target class.

\subsubsection{Normal Distribution}

Within this configuration, the number of samples in each class within each edge node follows a normal density function. Mathematically, this can be expressed as:

\begin{center}
$X \sim \mathcal{N}(\mu,\sigma^{2})$
\end{center}

where $\mu$ and $\sigma$ are defined as:

\begin{center}
$\mu = \frac{k \times N}{K}$, $\sigma = 0.2 \times N$
\end{center}

In the above equations, $k$ represents the ID of each edge node, $K$ denotes the total number of edge nodes, and $N$ corresponds to the total number of target classes in the training data. This configuration aims to provide varied distributions and different numbers of samples among different edge nodes, allowing each class to have a probability of having the majority of samples in a specific node.

\subsection{Machine Learning Method}

The implementation of the models in this study utilized Python and relied on the following libraries: torch 1.6.0 \cite{paszke2019pytorch}, torchvision 0.7.0 \cite{marcel2010torchvision}, and scikit-learn \cite{pedregosa2011scikit}, which were applied in model construction and evaluation.

To achieve satisfactory classification results, two distinct convolutional neural networks (CNN) \cite{krizhevsky2012imagenet} were trained for the MNIST and CIFAR-10 datasets. For the MNIST application, the CNN architecture comprised two 5x5 convolutional layers (with 10 output channels in the first layer and 20 in the second), each followed by 2x2 max pooling. Additionally, a fully connected layer with 50 units employing the ReLU activation function and a linear output layer were included.

For the CIFAR-10 application, the CNN architecture featured four 5x5 convolutional layers (with 66 output channels in the first layer, 128 in the second with a stride of 2, 192 in the third, and 256 in the fourth with a stride of 2). Furthermore, two fully connected layers utilizing the ReLU activation function, with 3000 and 1500 units respectively, were incorporated along with a linear output layer.

\subsection{Evaluation Metrics}
\label{sec:metric}

To assess the effectiveness of EdgeFL, three key metrics were selected: weights update latency, model evolution time, and model classification performance.

\subsubsection {Weights update latency}

Weights update latency measures the time it takes for the model to be transmitted. In centralized architectures which applied by existing FL platforms/frameworks, the central aggregation server collects the models. However, in the decentralized architecture of EdgeFL, where the aggregation function is moved to the edge, a peer node is ready to receive the updated model. The average weights update latency across all edge nodes during one training round is calculated. This metric provides insights into the network situation and communication overhead of each architecture option. Measurement of this metric involves checking the sending and receiving timestamps in all model receivers.

\subsubsection {Model Evolution time}

Model evolution time represents the time difference between two different versions of the deployed model at the edge nodes. Similar to weights update latency, the average model evolution time across all edge nodes during one training round is determined. This metric highlights the speed at which local edge devices update their knowledge, which is crucial for systems requiring quick adaptation to rapidly changing environments. Model evolution time is measured in all edge nodes by examining the model deployment timestamp.

\subsubsection{Model Classification Performance}

Model classification performance is a vital metric that indicates the quality of the trained model. It measures the percentage of correctly recognized images among the total number of testing images. The classification performance is evaluated on each edge device using their updated models. The test sample distribution should align with the training samples (local test set). The average classification performance across all edge nodes is reported.

\section{System Design of EdgeFL}

In this section, we present a comprehensive overview of the system design of EdgeFL. In addition, the APIs, functions and EdgeFL learning life-cycle are also presented in this section.

\subsection{System Design}

The EdgeFL allows for easy scalability, fault tolerance, and customization of the FL process. The framework consists of two main components: FL edge nodes and registration nodes. Edge nodes serve as independent participants, facilitating distributed and privacy-preserving model training without the need for a centralized server. The registration nodes act as coordination points to connect the FL edge nodes, enabling them to discover and communicate with each other in a decentralized manner.

\begin{itemize}
    \item FL Edge Nodes: The FL edge nodes are deployed on edge devices and play a crucial role in the FL process. Each FL edge node serves as a participant in the federated learning system. The FL edge nodes execute the FL training algorithm, exchange models with other nodes, and perform local model updates. The FL edge node code provided in the implementation utilizes the Flask framework to serve requested machine-learning model files from other peers.
    \item Registration Nodes: The registration nodes act as coordinators for the FL edge nodes. It maintains a list of active peers and provides services for registration, unregistration, and retrieval of peer information. The registration nodes enable FL client nodes to discover and communicate with each other. The implementation of the tracker server utilizes the Flask framework which exposes several APIs to facilitate the FL process.
\end{itemize}

\subsection{APIs and Services}

Table \ref{tab:apis} summarises the most important APIs and services for EdgeFL, including edge node registration and registration, peer information retrieval and model file serving.

\begin{table}[htbp]
\caption{APIs and services of EdgeFL}
\label{tab:apis}
\centering
\begin{tabular}{ccc}
\toprule
Name       & Endpoint & Method  \\ \midrule

Registration API     & /register   & POST                                 \\
Unregistration API & /unregister    & POST                         \\ 
Peer Information Retrieval API:    & /peers    & GET                            \\ 
Model File Serving API       & /latest\_model  & GET      \\ \bottomrule
\end{tabular}

\end{table}

\begin{itemize}
    \item Registration API: FL edge nodes send a registration request to the registration nodes through this API. The request includes the hostname of the FL edge node. Upon successful registration, the registration nodes add the peer information to their list of active peers.
    \item Unregistration API: FL edge nodes use this API to send an unregistration request to the registration nodes when they no longer participate in the FL process. The request includes the hostname of the FL edge node. The registration nodes remove the corresponding peer information from their list of active peers.
    \item Peer Information Retrieval API: FL edge nodes can query the registration nodes for a list of active peers using this API. The registration nodes respond with the list of active peer information, allowing FL edge nodes to discover and communicate with other peers.
    \item Model File Serving API: FL edge nodes expose this API to serve the requested model file. When a peer requests the latest model, the FL edge node responds by sending the machine-learning model file through HTTP response.
\end{itemize}

\subsection{Function Details and Example Usage}

The proposed EdgeFL framework allows software engineers to easily incorporate federated learning (FL) functionalities into their AI products. In contrast to complex FL platforms and frameworks, EdgeFL provides a streamlined implementation that requires only four lines of code (LOC). Because of this simplicity, software engineers can quickly integrate FL capabilities into their existing AI applications without requiring significant code changes or extensive re-engineering efforts. The following Listing 1 illustrates the example usage of the EdgeFL.

\begin{lstlisting}[language=Python, caption=Usage example of EdgeFL. , escapechar=@]
# --- Continue from node training part ---
# --- Initialize peer instance ---
@\textbf{peer = Peer(config)}@

# --- Start peer instance ---
@\textbf{peer.start()}@

for epoch in range(number_of_epochs):

    # --- Pull model from active peers and start aggregation
    @\textbf{w\_latest = peer.aggregation\_func()}@
    
    model.load_state_dict(w_latest)

    train(model, torch.device("cpu"), train_loader, optimizer, epoch)
    test(model, torch.device("cpu"), test_loader)
    scheduler.step()

    torch.save(model.state_dict(), "model-latest.pth")

# --- unregister from the registration node if leave
@\textbf{peer.unregister\_peer()}@

\end{lstlisting}

\textit{peer = Peer(configs)}: Initializing and creating an instance of the Peer class, representing a participant in the EdgeFL framework. The configs include addresses of registration nodes and the configuration of the customized aggregation function. This initialization step ensures that the FL edge node is properly configured to connect to the registration nodes and participate in the FL training and aggregation tasks. The peer object serves as a handle through which the FL edge node can interact with other peers, fetch models, register with the registration nodes, and perform aggregation operations.

\textit{peer.start()}: The function initiates the execution of the FL edge node within the EdgeFL framework. When invoked, this function triggers a series of actions that enable the FL edge node to participate in the FL process. It includes registering the FL edge node with the registration nodes, establishing connections with other peers and starting a background instance to serve asynchronous file requests from peers. By calling ``peer.start()'', the FL edge node becomes an active participant in the EdgeFL framework, contributing to the collaborative model learning while leveraging edge devices' capabilities.

\textit{peer.aggregation\_func()}: The function performs the aggregation process. When called, this function retrieves models from other FL edge nodes, as identified through the registration nodes, and applies the aggregation algorithm to combine these models into a single updated model. The aggregation function facilitates the collaborative nature of FL by leveraging the contributions of multiple peers to improve the overall model's accuracy and performance. By executing ``peer.aggregation\_func()'', the FL edge node actively contributes to the iterative model aggregation process, promoting the collective intelligence of the EdgeFL framework and enhancing the final model's quality.

\textit{peer.unregister\_peer()}: The function enables the FL edge node to gracefully exit from the EdgeFL framework. When invoked, this function notifies the registration nodes about the intention to unregister, providing the necessary information such as the hostname of the FL edge node. By calling ``peer.unregister\_peer()'', the FL client node initiates the process of removing itself from the active participant list maintained by the registration nodes. This action ensures the proper management of participants within the EdgeFL framework and allows for efficient resource allocation and coordination among the remaining active peers.

\subsection{EdgeFL Learning Life-Cycle}

\begin{figure*}[t]
\centerline{\includegraphics[scale=0.3]{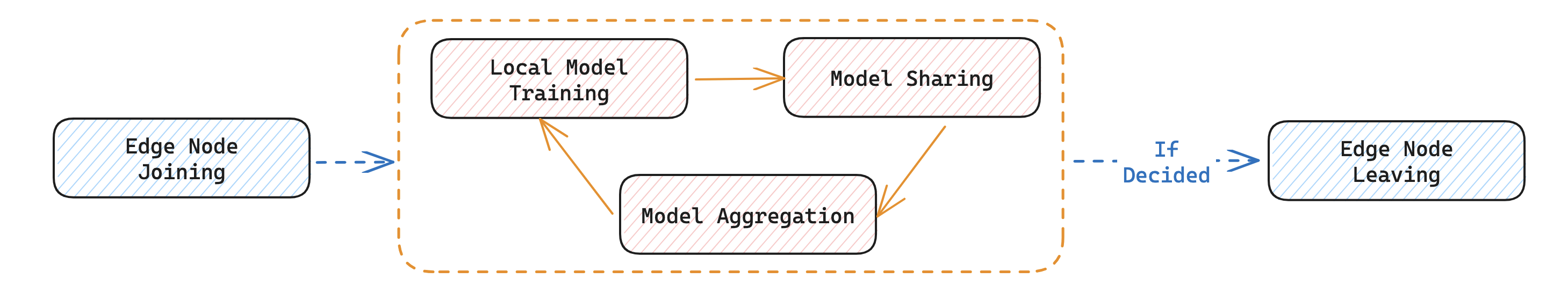}}

\caption{The Learning Life-Cycle of EdgeFL, including joining the FL process, model training, sharing, aggregation, and eventual node leaving. These stages collectively define the operational flow of EdgeFL within an edge node.}
\label{fig:fl-life}
\end{figure*}

The life cycle of the EdgeFL framework involves several key steps for an individual edge node to join, train, share models, aggregate, and eventually leave the FL process. Algorithm \ref{alg:edgefl} provides a detailed FL learning process of an individual edge node. The following is the description of each step:

\begin{enumerate}
    \item Edge Node Joining: The edge node initializes by creating an instance of the Peer class and background instance for model requests. The node then connects to the registration nodes, registers itself as an active participant, and obtains information about other peers.
    \item Model Training: The edge node starts the FL training process, performing local model training using its own dataset. It iteratively updates its local model to improve its performance. 
    \item Sharing Models: The FL client node retrieves models from other peers in the FL framework identified through the registration nodes. It fetches the latest models from other peers and incorporates them into its local model updates, benefiting from the knowledge and insights of other participants. It is worth noting that the model retrieval process occurs without disrupting the ongoing model training of the edge nodes, thanks to the background serving instance. This asynchronous model aggregation mechanism ensures uninterrupted model training while enabling the FL client node to actively contribute to the collaborative learning process.
    \item Model Aggregation: The FL client node executes an aggregation function, combining the locally updated models with the models obtained from other peers. The aggregation function integrates the diverse models to generate a new aggregated model that captures the collective knowledge of all participating nodes. It is important to highlight that the aggregation function in the EdgeFL framework can be customized to specific analysis and case requirements, providing software engineers with the flexibility to define and implement alternative aggregation functions that align with their specific needs. This paper utilizes a default averaging function for general performance analysis.
    \item Edge Node Leaving: When an edge node intends to leave the EdgeFL system, the FL client node notifies the registration server of its hostname for identification. The registration server updates the active participant list, removing the leaving edge node. However, it is worth noting that edge nodes have the option to remain in the system even after completing their learning process. By choosing to stay, these nodes contribute by providing their completed learning models to accommodate newly joined nodes. This approach ensures that the system benefits from the availability of finished-learning models, facilitating a seamless on-boarding experience for new participants in the EdgeFL framework.

\end{enumerate}

This life cycle repeats as new edge nodes join the FL framework, contribute to the training and aggregation processes, share their models, and eventually leave when they decide to end their participation. The EdgeFL allows for continuous collaborative learning and model improvement while maintaining the privacy and autonomy of individual edge nodes.

\begin{algorithm}[htpb]

    \caption{EdgeFL - In the system, $\alpha$ represents the ratio of aggregated peers; $C$ is the active peer list; B is the local mini-batch size; E represents the number of local epochs, and $\gamma$ is the learning rate.}
    \label{alg:edgefl}
    
    \begin{algorithmic}

	  \STATE Initialize $w_0$
        \STATE Initialize $\alpha$ as the ratio of aggregated peers
        \STATE Initialize $C$ as the active peer list
	
	\SetKwFunction{FMain}{Server\_Function} 
    \SetKwProg{Fn}{Function}{:}{}  
    \Fn{\FMain{}}{
        \FOR{each round t = 1, 2, ... } 

            \STATE Node\_Training(w\_t)

            \STATE Retrieval active peer list $C$
            
            \STATE$ m \longleftarrow max(len(C) \times \alpha, 1);$ 
            
            \STATE $N_t \longleftarrow $(random set of $m$ peers from $C$); 
            
            \FOR{each node $k \in N_t$ \textbf{in parallel}} 
            
            \STATE Fetch $w_{t+1}^{k}$ 
            
            \ENDFOR
            
            \STATE $ w_{t+1} \longleftarrow \sum_{k=1}^{K}\frac{1}{K}w_{t+1}^k;$ 
            
            \ENDFOR
    }

    \vspace{10pt}
    \SetKwFunction{FMain}{Node\_Training} 
    \SetKwProg{Fn}{Function}{:}{}  
    \Fn{\FMain{w}}{
        \STATE $ \beta \longleftarrow $(split $P_k$ into batches of size $B$);
        
        \FOR{each local epoch $i$ from 1 to E } 
            
            \FOR{batch $b \in \beta$} 
            
            \STATE $ w \longleftarrow w - \gamma\nabla l(w;b);$ 
            
            \ENDFOR
        
            \ENDFOR

        }
	    
        \STATE \textbf{return} $w$ to local storage

    \end{algorithmic}
   
\end{algorithm}

\subsection{Containerization and Scalable Deployment}

To ensure easy deployment on edge devices, the EdgeFL framework was containerized using containerization technologies, namely, Docker. The containerization process involved encapsulating all the necessary components, dependencies, and configurations of EdgeFL into a lightweight and portable container image. This approach allows for seamless deployment across a variety of edge devices, regardless of the underlying operating system or hardware architecture.

\begin{figure}[t]
\centerline{\includegraphics[scale=0.45]{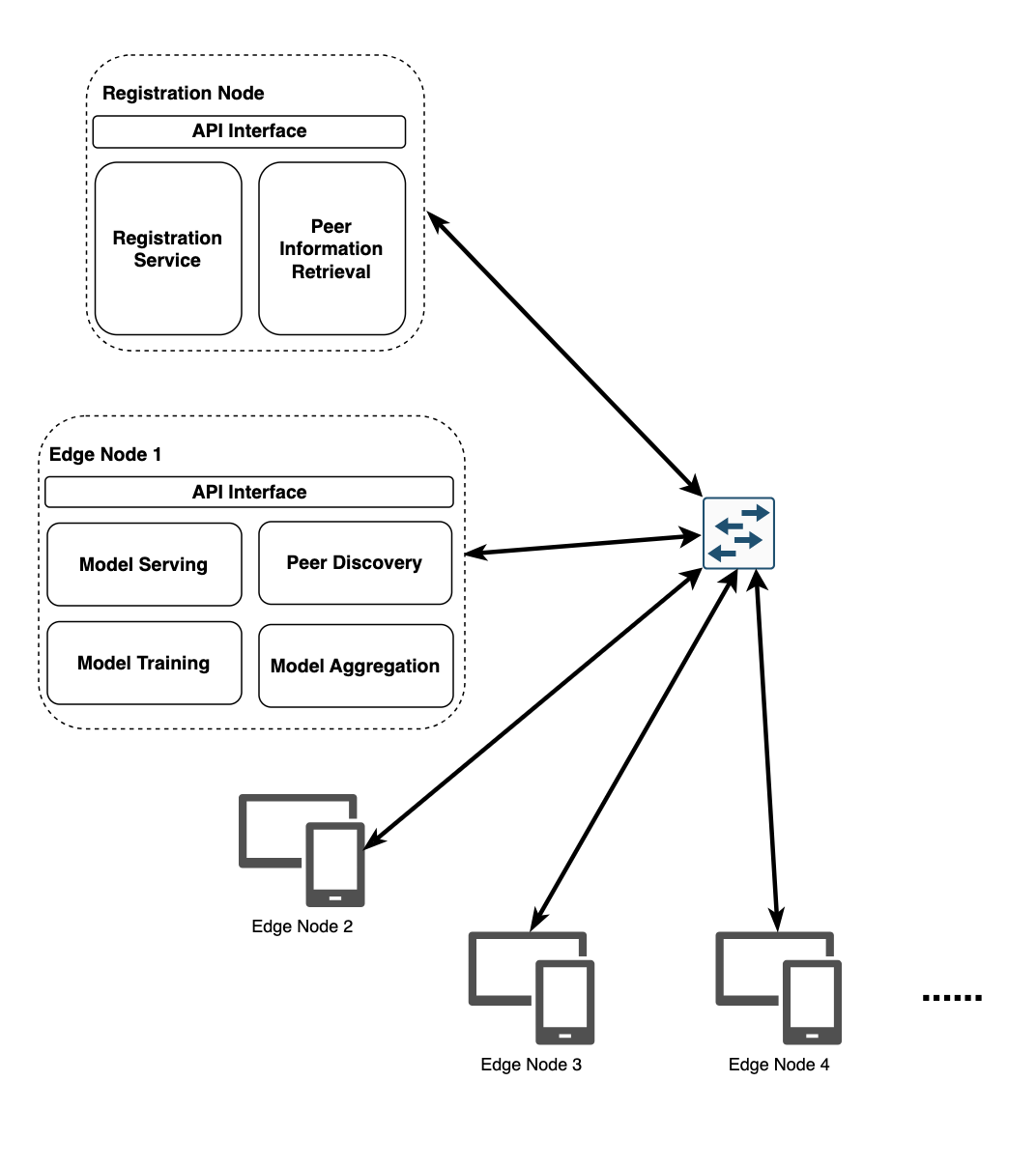}}

\caption{Containerized architecture for seamless and scalable deployment of the EdgeFL framework}
\label{fig:container}
\end{figure}

The architectural diagram illustrated in Figure \ref{fig:container} showcases the seamless and scalable deployment of the EdgeFL framework. Within this architecture, each edge node container includes services such as local model training, model aggregation, and model serving. Simultaneously, the registration node container contains services for registration and peer discovery. Through inter-connectivity, seamless communication is facilitated among all nodes within the framework. It is important to note that the number of registration nodes can be expanded in alignment with the number of participating edge nodes. This expansion enables efficient coordination and management within the EdgeFL framework, ensuring smooth coordination. 

By containerizing EdgeFL, software engineers can easily distribute and deploy the framework on edge devices without worrying about intricate installation procedures or compatibility issues. The containerized EdgeFL image contains all the required software libraries, frameworks, and configurations, providing a self-contained environment for running the FL client nodes. Additionally, containerization ensures that the EdgeFL framework remains isolated and independent, preventing conflicts with other software components on the edge device.

\section{Evaluation Results}

This section presents the experimental results of EdgeFL, focusing on three key aspects as defined in Section \ref{sec:metric}: (1) Weights update latency, which measures the time required to transmit model weights; (2) Model evolution time, which quantifies the duration for obtaining a new version of the model; and (3) Classification accuracy, evaluated on the edge dataset. To ensure an adequate number of samples on each edge node, our simulations involve a total of 10 nodes, with all nodes actively participating in the training procedure in both MNIST and CIFAR10 applications. This configuration enables comprehensive analysis and evaluation of the EdgeFL framework, providing valuable insights into its performance and effectiveness in real-world scenarios.

\begin{table*}[b]
\caption{Comparison of weight update latency and model evolution time by leveraging existing FL platforms/frameworks and our proposed EdgeFL framework}
\label{tab:time}
\begin{center}
\begin{tabular}{c c c  c c c }

\toprule
\multirow{2}{*}{FL Frameworks/Platforms} & \multicolumn{2}{c}{MNIST} & & \multicolumn{2}{c}{CIFAR10}\\ \cmidrule{2-3} \cmidrule{5-6}
 & \makecell{Weights Update Latency\\(sec)}  &  \makecell{Model Evolution Time\\(sec)} & & \makecell{Weights Update Latency\\(sec)}  &  \makecell{Model Evolution Time\\(sec)} \\

\midrule

TFF$^\star$ & - & 7.76 && - & 16.372\\
PySyft & 0.0247 & 9.05 && 0.0311 & 18.292\\
FATE & 0.0326 & 13.868 && 0.0473 & 31.689\\
LEAF$^\star$ & - & 10.239 && - & 27.362 \\
PaddelFL & 0.0258 & 11.667 && 0.0412 & 25.581\\

\textbf{EdgeFL} & \textbf{0.0092} & \textbf{5.093} && \textbf{0.0148} & \textbf{10.753}\\

\bottomrule
\multicolumn{6}{l}{\makecell[l]{$\star$ TFF and LEAF frameworks do not include actual server and client implementations but rather provide simulations of the FL process. \\Therefore, measuring weight latency is not feasible within these frameworks.}}\\

\end{tabular}
\end{center}
\end{table*}



First, we examine the weights update latency and model evolution time. Our experimental results (Table \ref{tab:time}) show that our proposed EdgeFL framework outperforms existing federated learning platforms/frameworks in terms of weights update delay and model evolution time in both MNIST and CIFAR10 applications. EdgeFL has smaller delays across the board when it comes to weights update delays. The reduced model delay indicates improved efficiency in transmitting models among edge nodes, demonstrating the effectiveness of our decentralized architecture. EdgeFL also excels at achieving rapid model evolution in scenarios with unequally distributed datasets. EdgeFL enables rapid model evolution by leveraging its decentralized architecture and effectively capitalizing on the pull-based model-sharing mechanism.  This mechanism allows edges to promptly update their local models, enhancing their knowledge based on the available data. Consequently, EdgeFL outperforms other frameworks by reducing the time required for model evolution in situations where dataset distribution is imbalanced.

These findings highlight the advantages of EdgeFL over traditional federated learning approaches. The superior performance in terms of model delay and evolution time can be attributed to the streamlined communication and aggregation processes within EdgeFL. By utilizing the power of edge computing, EdgeFL minimizes the network overhead and facilitates efficient model updates.

The observed improvements in model delay and evolution time have significant implications for real-world applications. The reduced delays enable faster transmission of updated models, ensuring timely access to the most recent knowledge across the edge network. Additionally, the shortened evolution time empowers edge devices to promptly adapt to evolving data features, making EdgeFL highly suitable for use cases that require quick model evolution and responsiveness to changing environments.

\begin{figure*}[t]
\centering
\subfigure[MNIST]{
\begin{minipage}[t]{0.48\textwidth}

\includegraphics[width=\textwidth]{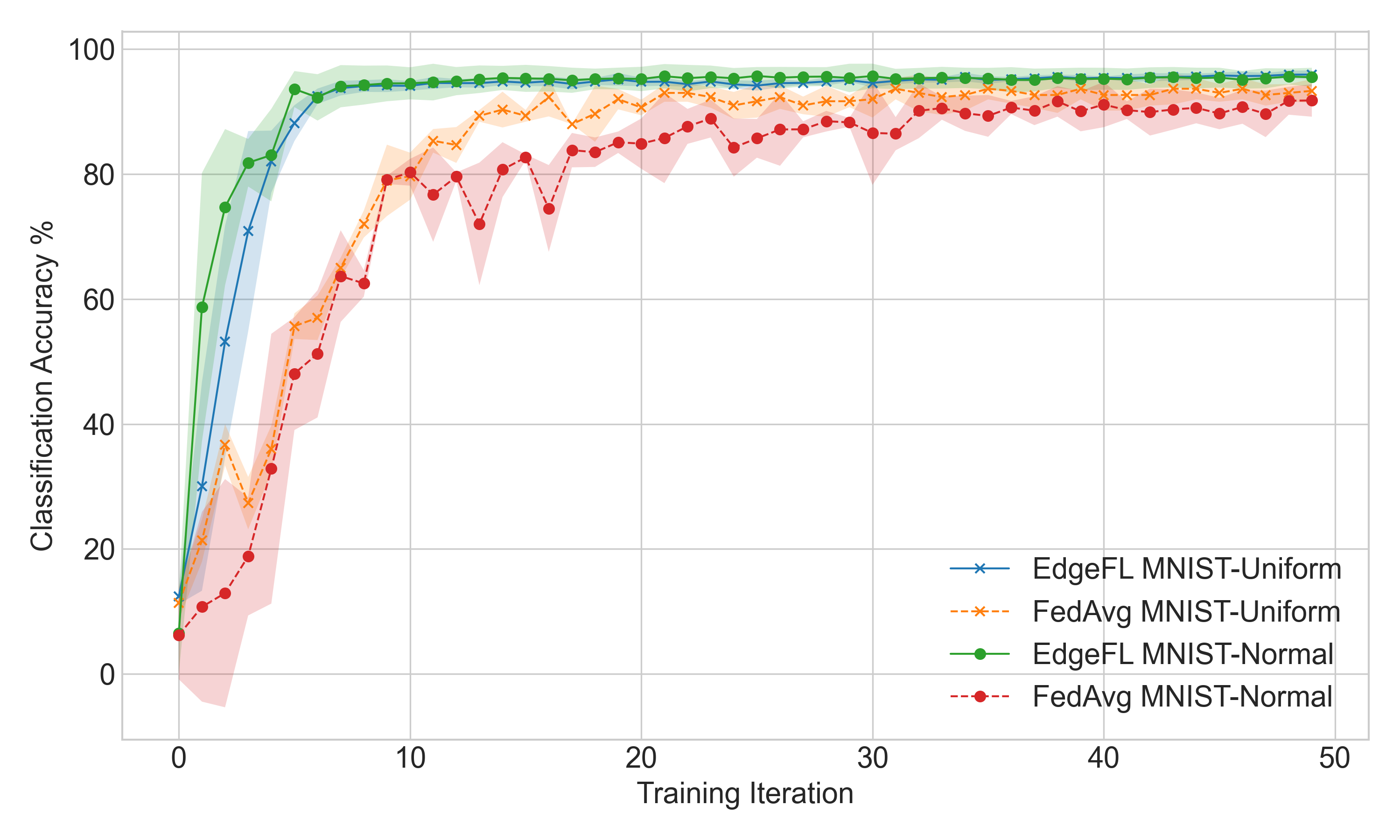} 

\label{fig:result-mnist}
\end{minipage}
}
\subfigure[CIFAR-10]{
\begin{minipage}[t]{0.48\textwidth}
\includegraphics[width=\textwidth]{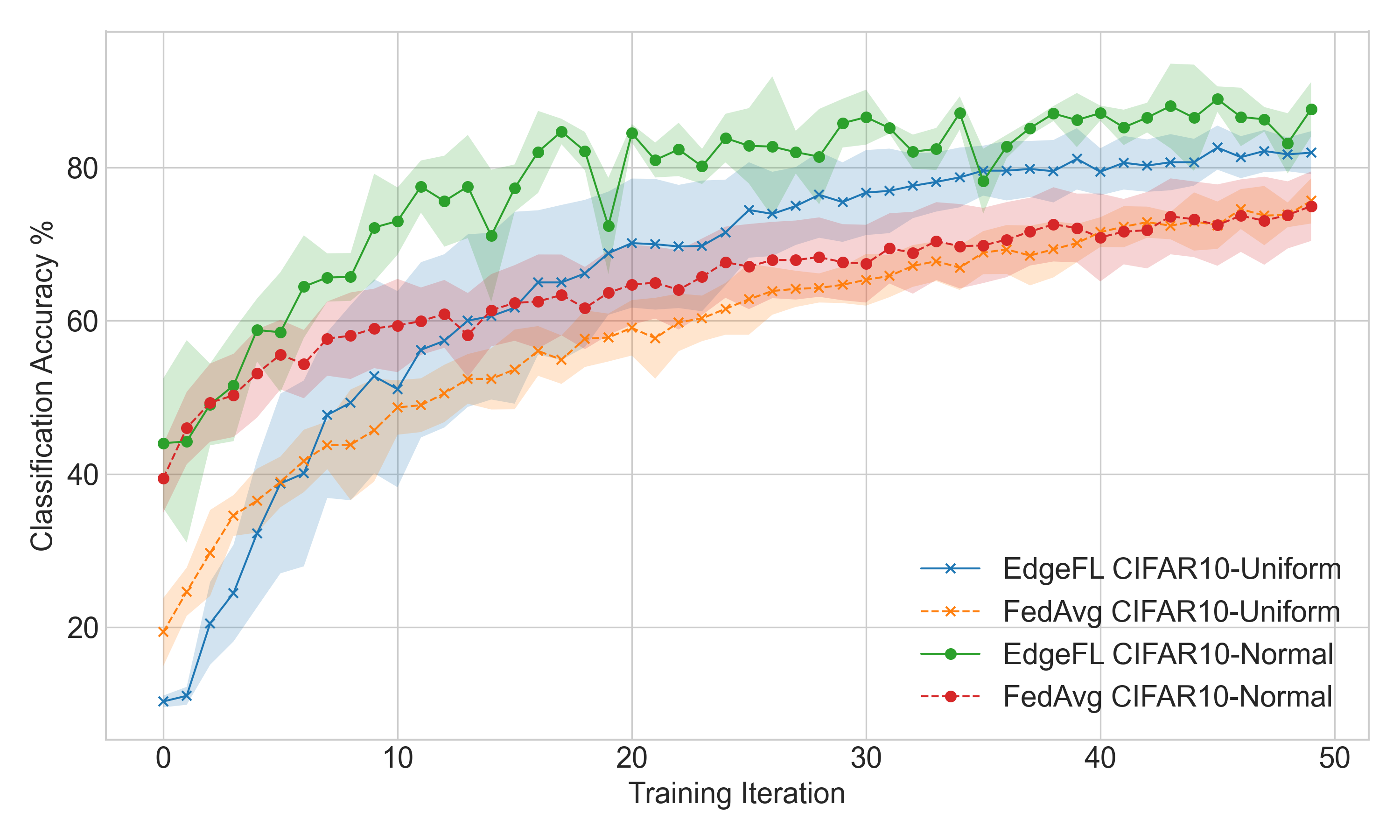}

\label{fig:result-cifar}
\end{minipage}
}

\caption{The comparison of classification accuracy by utilizing FedAvg (commonly used by existing FL platforms/frameworks) and  EdgeFL framework}

\label{fig:result}
\end{figure*}

In addition to evaluating weights update delay and model evolution time, we conducted extensive accuracy comparisons between EdgeFL's decentralized averaging and the widely used FedAvg algorithm \cite{mcmahan2017communication} found in existing federated learning platforms/frameworks. Figure \ref{fig:result} demonstrates the results, which reveal that decentralized averaging outperforms the centralized FedAvg approach when testing the models on edge devices. Notably, the average accuracy achieved by EdgeFL's decentralized averaging is approximately 2\% higher for MNIST and 5\% for CIFAR-10 datasets.

The observed increase in accuracy showcases the efficacy of EdgeFL's decentralized averaging mechanism in improving model performance. With the collective knowledge and insights from distributed edge devices, EdgeFL facilitates enhanced model convergence. As a result, EdgeFL enables more accurate and refined models, which are better equipped to handle the challenges of edge computing environments.

\begin{figure}[!htpb]
\centerline{\includegraphics[scale=0.35]{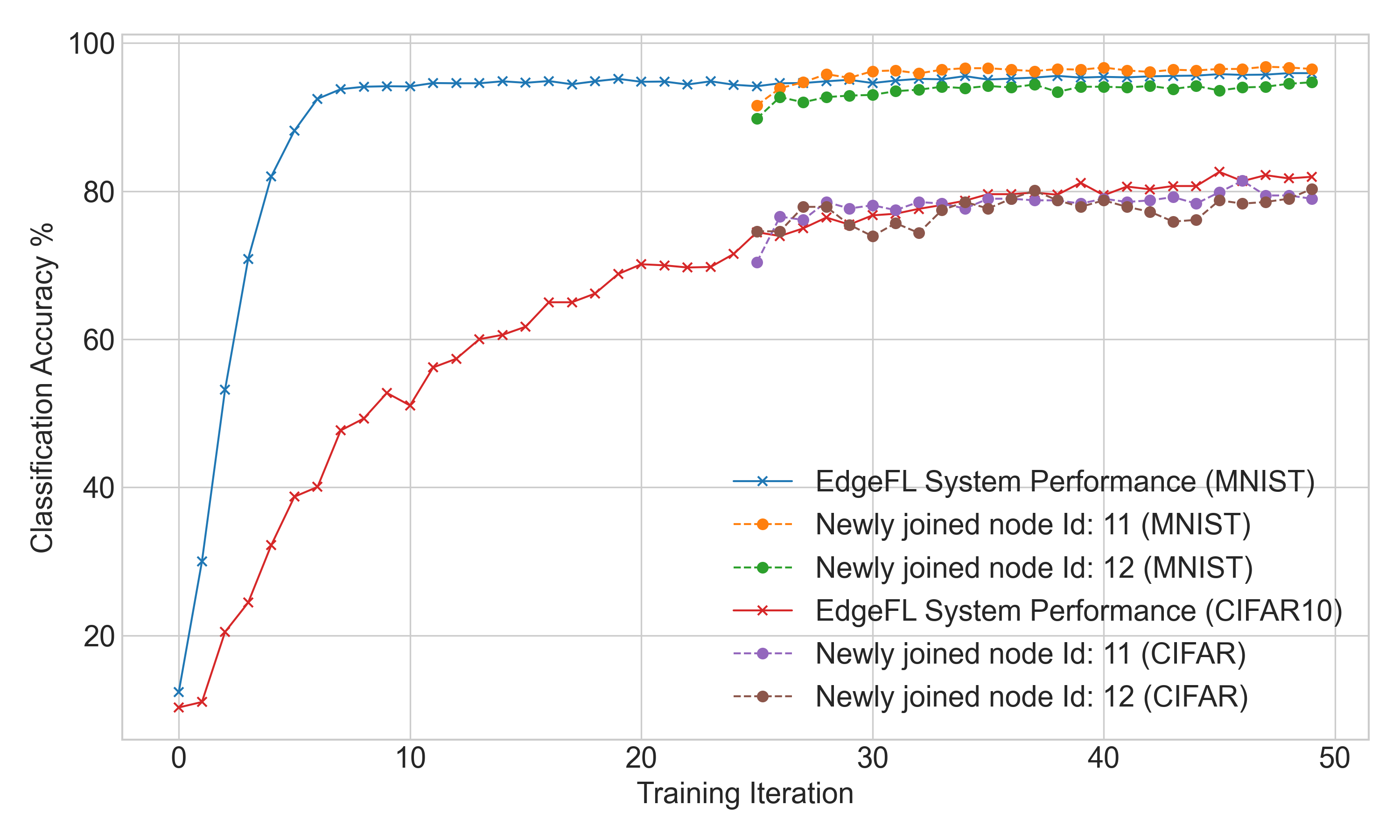}}

\caption{Midway joined node classification performance in both MNIST and CIFAR-10 application}
\label{fig:mifway}
\end{figure}

Furthermore, our study demonstrates the effectiveness of the asynchronous join feature in EdgeFL, which enables new nodes to seamlessly participate in the existing system and quickly acquire the latest knowledge without requiring retraining from scratch. As shown in Figure \ref{fig:mifway}, we observe that when new nodes (node id: 11, 12) join the system midway through the training process, they promptly attain the same accuracy level as the system accuracy. This outcome showcases the capability of EdgeFL to facilitate efficient knowledge transfer and rapid model convergence for newly joined nodes.

The asynchronous join functionality in EdgeFL offers significant advantages in terms of scalability and time-to-adaptability. By allowing new nodes to directly benefit from the collective intelligence of the system without the need for extensive training, EdgeFL significantly reduces the computational burden and time required for onboarding new participants. This feature is particularly valuable in dynamic environments where nodes frequently join and leave the system.

The demonstration of the asynchronous join capability in EdgeFL emphasizes its potential for real-world deployments, especially in scenarios where rapid knowledge transfer and quick integration of new nodes are crucial. By leveraging the existing knowledge base and facilitating seamless incorporation of new nodes, EdgeFL empowers federated learning systems to efficiently adapt and evolve over time. These findings highlight the practical benefits of EdgeFL's asynchronous join mechanism and its ability to enhance the scalability and flexibility of federated learning in dynamic edge computing environments.

\section{Discussion}

This paper focuses on analyzing and interpreting the results obtained from the experiments conducted with the EdgeFL framework. The evaluation of EdgeFL was performed using various metrics. Regarding weights update latency, EdgeFL demonstrated superior performance compared to existing Federated Learning platforms/frameworks. The decentralized communication mechanism employed in EdgeFL effectively reduced weights update delay by about 50\%, outperforming centralized alternatives. 

In terms of model evolution time, EdgeFL showcased notable advantages when dealing with unequally distributed datasets. The decentralized averaging approach facilitated faster model evolution compared to traditional Federated Learning methods, such as FedAvg. By utilizing the inherent pull-based mechanism, EdgeFL enables edge devices to swiftly acquire the latest knowledge from the system without the need to retrain from scratch. This capability is crucial for rapidly adapting to dynamically changing environments, making EdgeFL well-suited for real-world deployments.

The evaluation of classification accuracy revealed that EdgeFL's decentralized averaging mechanism consistently outperformed the centralized FedAvg algorithm when testing the model on edge devices. The observed increase of approximately 2\% and 5\% in average accuracy for MNIST and CIFAR-10 datasets demonstrates the effectiveness of EdgeFL in achieving better classification performance. Moreover, the framework enabled faster model convergence, contributing to improved overall efficiency and effectiveness.

Additionally, our experiments demonstrated the asynchronous join feature of EdgeFL, whereby a new node joining the system can promptly access the latest knowledge without requiring retraining from scratch. The experimental results showed that a newly joined node quickly achieved a high accuracy level, even when joining the system halfway through the training process. This feature highlights the scalability and efficiency of EdgeFL, as it enables seamless integration of new nodes into the existing network, without compromising overall performance.

Overall, the evaluation and analysis of the EdgeFL framework demonstrated its efficacy in addressing the challenges of scalable and efficient edge deployment in Federated Learning. The framework exhibited better performance in terms of weight update latency, model evolution time, and classification accuracy when compared to existing solutions. The decentralized averaging mechanism, along with its pull-based model-sharing approach, proved to be particularly advantageous for achieving faster model updates and improved convergence. These findings highlight the potential of EdgeFL for various real-world applications, particularly in industrial scenarios where timely and accurate model updates are critical.

\section{Conclusion}

In this paper, we presented EdgeFL, a novel edge-only decentralized federated learning framework that addresses the challenges of scalability, integration, and efficiency in edge deployments. By leveraging the edge-only model training and aggregation approach, EdgeFL eliminates the need for a central server, allowing for seamless scalability across diverse use cases. The framework offers a straightforward integration process, requiring only four lines of code (LOC) for software engineers to incorporate FL functionalities into their AI products. Additionally, EdgeFL provides engineers with the flexibility to customize aggregation functions according to their specific needs and requirements, enhancing the adaptability and versatility of the framework. Our experimental results and evaluation have highlighted the key strengths and advantages of EdgeFL. The framework outperforms existing FL platforms/frameworks in various aspects. It exhibits the capabilities of EdgeFL in reducing weight update latency and model evolution time by 50\% and improving classification accuracy by 2\% for the MNIST dataset and 5\% for the CIFAR dataset compared to existing FL platforms/frameworks. These findings emphasize the potential of EdgeFL in real-world applications, particularly in industrial scenarios where timely and accurate model updates are critical.

In future work, we will further validate and expand the capabilities of the proposed EdgeFL framework with more cases. We also intend to investigate resource optimization techniques such as model compression and quantization to enhance the communication efficiency of edge devices in EdgeFL. Furthermore, the adaptive aggregation strategies that dynamically adjust the aggregation process based on network conditions, device capabilities, and data heterogeneity will also be explored.

\bibliographystyle{IEEEtran}
\bibliography{cite}

\end{document}